# Hidden hydroxides in KOH-grown BaNiO$_3$ crystals: a potential link to their catalytic behavior

Lun Jin[1, *], Haozhe Wang[2], Xianghan Xu[1], Danrui Ni[1], Chen Yang[1], Yu-Chieh Ku[3], Cheng-En Liu[3], Chang-Yang Kuo[3,4], Chun-Fu Chang[5], Raimundas Sereika[6], Wenli Bi[6], Weiwei Xie[2], and Robert J. Cava[1, *]

[1] Department of Chemistry, Princeton University, Princeton, New Jersey 08544, USA

[2] Department of Chemistry, Michigan State University, East Lansing, Michigan 48824, USA

[3] Department of Electrophysics, National Yang Ming Chiao Tung University, Hsinchu 30010, Taiwan

[4] National Synchrotron Radiation Research Center, Hsinchu 30076, Taiwan

[5] Max Planck Institute for Chemical Physics of Solids, Nöthnitzer Strasse 40, 01187 Dresden, Germany

[6] Department of Physics, University of Alabama, Birmingham, Alabama 35294, USA

* E-mails of corresponding authors: ljin@princeton.edu; rcava@princeton.edu

**Abstract:** The hexagonal perovskite BaNiO$_3$, prepared via non-ceramic approaches, is known to act as a good catalyst for the oxygen-evolution reaction (OER) in alkaline media. Here we report our observation that BaNiO$_3$ synthesized via KOH flux growth and high O$_2$ pressure ceramic synthesis have different magnetic properties. We show that this is because the KOH flux-grown crystals made in open-air are actually a hydroxide-containing form of BaNiO$_3$ that can be "dried" upon annealing in O$_2$ flow. This work not only unveils a previously unknown aspect of the BaNiO$_3$ OER catalyst and offers some insights into the underlying mechanism, but also suggests that hydroxide ions may be present in other hexagonal perovskite oxides prepared in wet conditions.

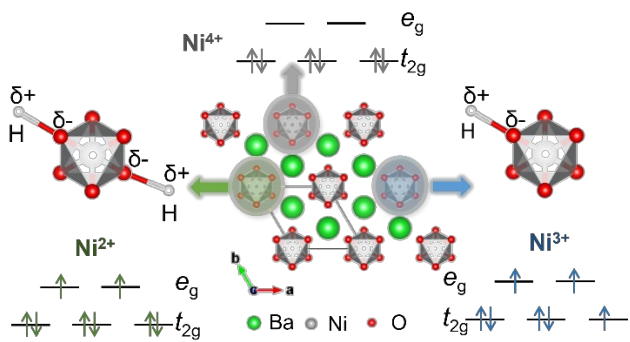

**Introduction**

Complex metal oxides exhibit enriched structural chemistry and sometimes unexpected magnetic and electronic properties, ranging from magnetism, superconductivity, and magnetoresistance to ionic conductivity and unusual catalytic behavior.[1–8] The $ABO_3$ perovskites, as one of the largest oxide structural families, serve an important role in this context. In standard perovskites, which are based on corner-shared $BO_6$ octahedra, the metal centers are relatively far from each other. In contrast, $ABO_3$ materials based on face-shared $BO_6$ octahedra, as are found in hexagonal oxide perovskites, are more rare and display significantly smaller B-O-B bond angles and shorter distances between adjacent metal centers. Therefore face-sharing octahedra can have a dramatic impact on the magnetic properties, and thus materials such as hexagonal perovskites have drawn enduring attention from the science community.[9]

$BaNiO_3$ is an ideal 2H hexagonal perovskite where what are formally $Ni^{4+}O_6$ octahedra share faces and form infinite 1D chains along the hexagonal *c*-axis. This material was initially reported as being prepared via high $O_2$ pressure solid-state synthesis[10,11]. Later, other synthetic approaches such as molten salt flux[12,13], sol-gel[13,14] and thermal decomposition[15] were also employed to make it. $BaNiO_3$ prepared via high $O_2$ pressure solid-state synthesis was reported to be diamagnetic and has been widely accepted as such[11,16,17]. Researchers have been working on several oxygen-deficient analogs of the $BaNiO_{3-\delta}$ type and it is found that the structural and magnetic properties are highly dependent on the concentration of oxygen vacancies in the lattice[18–23].

In recent years, $BaNiO_3$ prepared via non-ceramic approaches has been found to act as a highly functional catalyst for the oxygen-evolution reaction (OER) in alkaline media, which is an indispensable process for diverse energy storage devices.[13,14] The discovery of $BaNiO_3$ as a cost-effective and highly active catalyst towards the OER garners it renewed attention in the chemistry community. Similar catalytic behavior has also been reported for $ACoO_3$ (A = Ca, Sr).[24] However, recent studies on $BaNiO_3$ prepared via these non-ceramic approaches have mainly focused on its catalytic behavior, while its fundamental structural and physical properties have generally been overlooked.

In the present work, we have grown "$BaNiO_3$" crystals using a molten KOH flux. However, the as-grown crystals show an unexpected magnetic phase transition at low temperature, in contradiction to what is seen for $BaNiO_3$ prepared via conventional high $O_2$ pressure solid-state synthesis[11]. Upon further oxygen annealing, the magnetic transition vanishes, and the crystals are diamagnetic. Although the magnetic properties are consistent with those reported for BaNiO$_{3-\delta}$, no oxygen vacancies were detected in the as-grown crystals. To shed light on this issue, we have performed a series of analyses and found that the KOH flux-grown crystals made in open-air are indeed a "wet" form of BaNiO$_3$, unlike the material made in a relatively "dry" synthetic condition under high O$_2$ pressure that stabilizes the unusually high oxidation state of Ni$^{4+}$. We find that a significant portion of the O$^{2-}$ ions in the anion-lattice are actually in the form of OH$^-$, and hence necessarily reduce some of the Ni$^{4+}$ centers into Ni$^{3+}$ or even Ni$^{2+}$ in a disordered manner. Although we do not rule out the possibility that the OH$^-$ content of crystals may vary upon different reaction conditions or crystal-growth media, the conditions adopted in the present study yield an average composition of BaNiO$_{2.45}$(OH)$_{0.55}$, which we detect through measuring the average valence state of Ni (+3.45) in the KOH flux-grown crystals determined by X-ray absorption spectroscopy. This information not only unveils a previously unknown aspect of the OER catalyst BaNiO$_3$ and offers some insights into the underlying mechanism — filling electrons into the originally empty $e_g$ orbital of Ni$^{4+}$ is considered a descriptor for OER activity[25], but also suggests that hidden hydroxide ions may also be present in other hexagonal perovskites prepared in wet conditions.

**Experimental**

The as-grown dark golden-green needle-shaped crystals were synthesized via a flux-mediated crystal growth method. (1) A stoichiometric ratio (1:1) of starting reagents BaCO$_3$ (Alfa Aesar, 99.997%) and NiO (Alfa Aesar, 99%) was thoroughly ground together using an agate mortar and pestle, and then transferred into an alumina crucible. The reaction mixture was first slowly (1 °C/min) heated to 1000 °C in air and held overnight to decompose the carbonate, and then annealed in air at 1100 °C (3 °C/min) for 24 hours with intermittent grindings. The reaction progress was monitored using laboratory X-ray powder diffraction data collected at room temperature on a Bruker D8 FOCUS diffractometer (Cu Kα) over the 2θ range between 5° and 70°. The main phase of the dark green powder precursor was found to be Ba$_3$Ni$_3$O$_8$. (2) 0.5 g powder precursor was then placed in an alumina crucible together with 0.5 g Ba(OH)$_2$·8H$_2$O (Alfa Aesar, 97%) and 50 g KOH (Alfa Aesar, 99.98%). The filled crucible was capped with a loosely-fitting alumina lid and then heated to 700 ˚C (3 °C/min) in air for 24 hours, followed by slow cooling down (18 °C/h) to room temperature. (3) The residue KOH flux was then slowly dissolved by adding DI water, and the reaction mixture was vacuum-filtered a couple of times to isolate the dark golden-green needle-shaped crystals. Ba(OH)$_2$·8H$_2$O was used to suppress the formation of NiO during crystal growth, as suggested in previous studies[12,13]. A portion of the as-grown crystals was annealed in O$_2$ flow for 24 hours at 600 ˚C.

The powder precursor prepared in (1) was also used for a high-pressure solid-state synthesis of BaNiO$_3$ powder. A 1:1 volume ratio mixture of the precursor and KClO$_3$ powder was packed (separately) into a sealed gold crucible. A gold spacer was used to separate the two materials and to facilitate the isolation of the target phase. The gold crucible was then packed into a boron nitride crucible-and-lid set. The set was pressed to 4 GPa in a cubic multi-anvil system (Rockland Research Corporation) and heated at 800 ˚C for 200 minutes. The system was quench-cooled before depressurizing. The resulting product was black and polycrystalline. The X-ray pattern indicated the full decomposition of the KClO$_3$ and the formation of the target hexagonal compound.

The crystal structure of an as-grown crystal was determined by single crystal X-ray diffraction. The as-grown crystal with dimensions 0.444 × 0.102 × 0.091 mm$^3$ was mounted on a nylon loop with paratone oil, and characterized using a XtalLAB Synergy, Dualflex, Hypix single crystal X-ray diffractometer with Mo K$_\alpha$ radiation ($\lambda$ = 0.71073 Å, micro-focus sealed X-ray tube, 50 kV, 1 mA) equipped with an Oxford Cryosystems low-temperature device, operating at $T$ = 100.0(2) K. CRI Data were measured using $\omega$ scans. The total number of runs and images was based on the strategy calculation from the program CrysAlisPro 1.171.42.79a (Rigaku OD, 2022). Data reduction was performed with correction for Lorentz polarization. The integration of the data using a hexagonal unit cell yielded a total of 10154 reflections to a maximum $\theta$ angle of 48.16° (0.48 Å resolution), of which 274 were independent (average redundancy 37.058, completeness = 100.0%, R$_{int}$ = 6.71%). Empirical absorption correction using spherical harmonics, implemented in the SCALE3 ABSPACK scaling algorithm was employed. The structure was solved and refined using the SHELXTL Software Package, in the centrosymmetric space group $P6_3/mmc$. The crystallographic data are listed in the Supporting Information.

Laboratory powder X-ray diffraction data with good statistical significance, covering a 2$\theta$ range between 5° and 110°, were collected from crushed and ground as-grown crystals; the bulk refinement of these data was carried out using the GSAS-II program. Scanning electron microscopy (SEM) images of as-grown crystals were collected using a Quanta 200 FEG Environmental-SEM. Thermogravimetric Analysis (TGA) was conducted using a TA Instruments

TGA 5500 under steady flowing $O_2$ gas. Approximately 4 mg of as-grown crystals were loaded into a platinum pan and heated from room temperature to 600 °C at a constant rate of 10 °C/min. The system was kept under flowing $O_2$ at 600 °C for 2 hours, and then fast cooled to room temperature. The post-TGA sample was tested by lab p-XRD and magnetic susceptibility measurements.

Diffuse-reflectance (DR) experiments were performed at ambient temperature on a Cary 5000i UV–vis–near-IR spectrometer equipped with an internal DRA-2500 integrating sphere. The reflectance data were transferred to pseudo-absorbance using Kubelka–Munk theory. Pre-dried pure KBr was used for the baseline correction. The Raman spectra were collected on a single crystal sample using the Raman Lab System of the GSECARS[26], Advanced Photon Source, Argonne National Laboratory with 532 nm laser excitation. A 4:1 methanol-ethanol mixture was used as the pressure transmitting medium (PTM) and the pressure in the diamond anvil cell (DAC) was monitored by the R1 fluorescence line of ruby up to 32.4 GPa.[27]

X-ray absorption spectroscopy (XAS) was performed at the TPS 45A1 beamline of the Taiwan Photon Source. The Ni $L_{2,3}$ XAS spectra were taken in the total electron yield mode with a photon energy resolution of 0.1 eV. A single crystal NiO was measured simultaneously for energy calibration. Three configurations: $3d^6$, $3d^7\underline{L}^1$, and $3d^8\underline{L}^2$ for the ground state and $2p^53d^7$, $2p^53d^8\underline{L}^1$, and $2p^53d^9\underline{L}^2$ for XAS excited state were employed in the calculations to fit these data, where $\underline{L}^n$ denotes the number of holes in the ligand shells that hybridize with the Ni $3d$ orbital. $pd\sigma = 1.74$eV and $pd\pi = 0.74$eV were used to define the hopping strength between Ni $3d$ and the ligand orbitals; 10Dq=1.0eV is used to define the octahedral crystal field. The charge transfer energy $\Delta = -4.0$eV and Coulomb repulsion $U_{dd} = 6.0$eV and $U_{dp} = 7.5$eV were used to define the central energy between each configuration. The Slater Integral of $F_{dd}^2$, $F_{dd}^4$ and $F_{dp}^2$, $G_{dp}^1$, $G_{dp}^3$ reduced to 65% from Hatree–Fock value were also employed in the calculations to fully implement the atomic-like multiplet effect that strongly shapes the XAS.

The magnetization data were collected using the VSM option of a Quantum Design Physical Property Measurement System (PPMS). Temperature-dependent magnetization (*M*) data were collected under an applied external field (*H*) of 1000 Oe. Magnetic susceptibility is defined as M/H. Field-dependent isothermal magnetization data between *H* = 90000 Oe and –90000 Oe were collected at *T* = 300 K and 1.8 K. Heat capacity was measured using a standard relaxation method in the PPMS over the temperature range 0.5 to 20 K.

**Results and Discussion**

The crystal structure of one of the needle-shaped crystals grown from the KOH flux is shown in **Figure 1a**. The sample crystallizes, in good agreement with previous reports of BaNiO$_3$[12–14,28], in the hexagonal space group $P6_3/mmc$ (No. 194). The lattice parameters refined from room temperature powder XRD data collected on all prepared materials in the present work are nearly undistinguishable from those reported in literature ($a$ = 5.64 Å, $c$ = 4.81 Å)[12–14,28], while our single crystal XRD data are collected at 100 K, hence a slight contraction of the unit cell ($a$ = 5.62 Å, $c$ = 4.79 Å) is expected (**Table S1**). In the crystal structure, the face-sharing NiO$_6$ octahedra form parallel one-dimensional chains separated by Ba atoms. The intrachain Ni-Ni distance is approximately 2.40 Å. Between chains, the Ni-Ni distance is about 5.62 Å. The 6$h$ oxygen site occupancy was carefully examined in the refinement, but we do not observe any vacancies. The atomic coordinates and equivalent isotropic atomic displacement parameters determined are shown in **Table S2**.

To make sure that all the properties measured in the present work are reliable, the bulk material was examined by powder X-ray diffraction. The needle-shaped crystals tend to show strong preferred orientation, which in principle would cause intensity mismatch for peaks in the powder X-ray diffraction pattern. Thus, the as-grown crystals were crushed and ground thoroughly to minimize the effects of preferred orientation. The major peaks can all be indexed by the hexagonal unit cell of BaNiO$_3$ (space group $P6_3/mmc$), consistent with the SC-XRD results, while a few low intensity peaks that are unaccounted for may belong to the trace amount of residue flux left in the bulk material (**Figure 1b**). The as-grown needle-shaped crystals genuinely show a hexagonal cross section with a diameter around 10 microns, while their lengths range from 0.1 to 1 millimeter (**Figure 1c**).

The Raman spectra up to 32.4 GPa were collected to examine the robustness of the lattice of the as-grown crystals and provide data for others who may be interested in the characterization of nanoparticles (**Figure 2a-b**). Three Raman active modes, at about 415 (M1), 503 (M2) and 597 (M3) cm$^{-1}$, are observed at ambient pressure, consistent with the Raman active modes of a hexagonal $P6_3/mmc$ structure material with the 2$a$, 2$d$, and 6$h$ sites occupied (A$_{1g}$, E$_{2g}$, and E$_{1g}$). Within the frequency range studied in the Raman spectra, no indication of structural anomalies was observed, with all three modes moving smoothly to higher frequencies under pressure. On pressure release back to about 3.5 GPa, there the three modes remain, showing that the crystallinity of the material has remained intact. The evolution of the Raman active modes under pressurization is summarized in **Figure 2c**. A slope change in fitted black lines occurs at ~12 GPa. This subtle change may either be associated with an electronic transition in $BaNiO_3$ or arise from a glass transition in the pressure transmitting medium, but further work would be needed to clarify this issue.

The band gap for the as-grown crystals was determined from the diffuse reflectance spectra. The pseudo-absorbance, transferred from reflectance using the Kubelka-Munk function, is plotted against wavelength (nm) in **Figure S1**. The optical transition was then analyzed based on Tauc plots. The extracted value for the band gap is approximately 1.5 eV if the direct-transition equation is applied or 1.1 eV if the indirect-transition approach is adopted. The relatively small magnitude of band gap observed is consistent with the sample's dark color, as well as the non-metallic behavior of KOH flux-grown crystals.

The KOH flux-grown crystals of $BaNiO_3$ thus initially look to consist solely of $Ni^{4+}$ as expected, however an intriguing phenomenon was observed during our magnetic property characterization. The crystals were found to show a magnetic moment and a clear magnetic transition in the low-temperature regime, i.e., at least a portion of the Ni cations in the lattice are magnetic. This is unexpected since $BaNiO_3$ should be diamagnetic, because $Ni^{4+}$ adopts a low-spin $d^6$ configuration with no unpaired electrons. The first-instinct explanation for this circumstance is that oxygen vacancies are present in the as-grown crystals, however, that would contradict the SC-XRD results in which no oxygen vacancies were detected. The TGA results further confirm the SC-XRD results. The collected data clearly demonstrate a decrease in the sample's mass upon heating in a steady stream of $O_2$, instead of an increase in mass which would be expected for filling the oxygen vacancies present in the lattice, if any (**Figure S2**). We postulate that the ~ 0.25% mass loss (step heights corrected from the drift) in the TGA under an oxidative condition is due to the "drying" process, which may transform the $OH^-$ ions in the lattice back to $O^{2-}$, hence oxidize the $Ni^{2+}$ and $Ni^{3+}$ to $Ni^{4+}$, accompanied by the formation of $H_2O$ which evaporates upon heating. The amount of $OH^-$ ions calculated from the TGA result matches well with the average formula based on the XAS results. What's more, the fraction of oxygen vacancies needed to cause the onset of magnetic ordering between magnetic Ni centers in this material should be non-trivial and thus easily detected, but their presence is not alerted by any of the structural or composition characterization techniques that we have employed in the present work. What's more, after annealing the KOH flux-grown crystals in the $O_2$ flow, the previously observed magnetic transition vanishes, and the $O_2$-annealed crystals genuinely behave diamagnetically in the applied external magnetic field, consistent with $BaNiO_3$ prepared via high $O_2$ pressure ceramic synthesis.

In order to map out the actual valance state of Ni in our KOH flux-grown crystals and $O_2$-annealed crystals, we have employed X-ray absorption spectroscopy (XAS), a powerful tool for determining the valance state of $3d$ transition metal elements, since the energy position and multiplet spectral features at the $L_{2,3}$-edge are highly sensitive to the valence state and local environment.[29,30] We measured the Ni $L_{2,3}$-edge XAS (**Figure 3**), which shows a shift to higher energy with increase in Ni valence state, from NiO (purple) to $LaNiO_3$ (red)[31] and further to our crystals (blue). The energy difference originates from the different Coulomb interaction between the 2p core hole and different numbers of valence electrons. To understand the experimental Ni-$L_{2,3}$ XAS spectrum and hence to accurately determine the Ni valence state in both the as-grown and $O_2$-annealed crystals, we calculate the Ni-$L_{2,3}$ XAS spectrum of $Ni^{4+}$ using a configurational interaction cluster calculation, considering that the theoretical XAS spectrum of $Ni^{4+}$ is rarely documented[32]. The theoretical approach employed considers full multiplet and local solid effects as well as the $3d$-$O2p$ covalence. The calculated Ni-$L_{2,3}$ XAS spectrum of $Ni^{4+}$ is shown in bottom (green) of each panel in **Figure 3**. By manipulating the ratio between Ni with different valences, it can be seen that for both as-grown and $O_2$-annealed crystals, the sum of the calculated XAS of $Ni^{4+}$ and the experimental XAS of $LaNiO_3$/NiO can nicely reproduce the experimental XAS collected from our crystals. For as-grown crystals (**Figure 3a**), the spectral weight ratio indicates $Ni^{2+}$: $Ni^{3+}$: $Ni^{4+}$ = 0.15: 0.25: 0.60, corresponding to an average $Ni^{3.45+}$ valence state. In contrast, for $O_2$-annealed crystals (**Figure 3b**), the spectral weight ratio indicates $Ni^{3+}$: $Ni^{4+}$ = 0.3: 0.7, corresponding to an average $Ni^{3.7+}$ valence state. (There is one thing to note for the $O_2$-annealed crystals - all synthetic-lab-based characterizations described in the present work were performed on "freshly" annealed crystals, i.e., sample degradation can be neglected. However, it is impractical for us to protect the $O_2$-annealed crystals from sample degradation ($Ni^{4+}$ tends to slowly reduce if the $O_2$ partial pressure is not high enough, such as in air) for the characterizations carried out in the XAS facility. Therefore, it is not surprising that a small portion of $Ni^{3+}$ is found in $O_2$-annealed crystals since they have been exposed to air for more than a month. The diagnostic feature here is that the $O_2$-annealed crystals, even after a period of degradation, still have a much higher average Ni valence state ($Ni^{3.7+}$) than the as-grown crystals ($Ni^{3.45+}$).

A detailed magnetic study was performed on the as-grown crystals and $O_2$-annealed crystals. $BaNiO_3$ powder prepared via the high-pressure method was used as a reference. The magnetic susceptibility $\chi$ (defined as M/H) for each material is plotted against temperature T (**Figure 4a**). For KOH flux-grown crystals, a divergence between the zero-field-cooled (ZFC, black) and field-cooled (FC, red) curves is clearly observed at T ≈ 15 K after going through a local maximum in the magnetic susceptibility. The DC magnetic susceptibility data, over the suitable temperature range (selected as the straight-line part of the green $1/\chi$ vs. T curve, marked in orange) were fitted to the Curie-Weiss law ($\chi = C/(T - \theta) + \chi_0$), to yield the Curie constant $C = 0.0711(2)$ $cm^3$ K $mol^{-1}$, and hence indicating an observed $\mu_{eff}$ value of 0.75 $\mu_B$/f.u. This corresponds to ~ 54% of the spin-only value (1.39 $\mu_B$/f.u.) deduced from a linear combination of 15% $S = 1$ $Ni^{2+}$, 25% of $S = 3/2$ $Ni^{3+}$ and 60% of $S = 0$ $Ni^{4+}$ centers suggested by the XAS data. The Weiss temperature $\theta$ is +16.3(4) K which indicates that the dominant magnetic coupling is ferromagnetic. In contrast, the magnetic susceptibilities of the $O_2$-annealed crystals (**Figure 4**, blue) and the $BaNiO_3$ powder prepared via high-pressure ceramic synthesis (**Figure 4**, magenta) are genuinely temperature-independent and have dramatically smaller magnitudes when compared to the as-grown crystals, with a small tail at low temperature which is common for diamagnetic materials, attributed to the presence of a small % of spins that are not coupled to the main magnetic system (less than 1 % in our case).

The field-dependent magnetization data were collected from each phase, at T = 300 K and 1.8 K, and are plotted as M against H ($-9 \leq H / T \leq 9$) in **Figure 4b**. In general, the 300 K isotherms collected as a function of applied field are linear in the whole applied field range and pass through the origin, with a positive slope for as-grown crystals and a negative slope for $O_2$-annealed crystals and $BaNiO_3$ powder prepared via high-pressure, respectively. The analogous data collected at 1.8 K show a significant enhancement in magnitude for the as-grown crystals with a subtly opened hysteresis loop, all consistent with the magnetic susceptibility data. In addition, to explore the degradation of the $O_2$-annealed crystals in air, a batch of "freshly" $O_2$-annealed material was deliberately left in air for 5 days and hence more than a month (mimicking the XAS sample), and magnetic data were re-collected (**Figure 4c**). It is found that the magnetic susceptibility sequentially grows in magnitude with respect of the air-exposure span, but not to the scale observed for as-grown crystals, and the magnetic transition is still absent, consistent with the XAS results and with our proposal that $Ni^{4+}$ tends to slowly reduce when the $O_2$ partial pressure in the ambient is not high enough.

Specific heat capacity data were collected from as-grown crystals over the temperature range $1.8 \leq T / K \leq 20$ in various external applied fields ($\mu_0 H$ = 0, 3, 6, 9 T). Analogous data were also collected for non-magnetic $O_2$-annealed crystals (**Figure S3**). The total heat capacity of a magnetic material can be interpreted as the sum of electronic, phonon and magnetic contributions ($C_{total} = C_{electron} + C_{phonon} + C_{mag}$). Thus to isolate the magnetic contribution $C_{mag}$ to determine the magnetic entropy change $\Delta S_{mag}$, the heat capacity contributions due to electrons ($C_{electron}$) and phonons ($C_{phonon}$) should be estimated and subtracted. To make this approximation, the heat capacity data collected from non-magnetic freshly $O_2$-annealed crystals serves as an approximation for $C_{phonon}$ of as-grown crystals. $C_{electron}$ is effectively = 0 for non-metallic materials such as these (the resistivity of all the materials prepared in this study is too high to measure, meaning that there are no conduction electrons present that would yield a significant $C_{electron}$). The resulting $C_p/T$ is plotted against temperature in **Figure 5a** (the total heat capacity $C_{total}$ data are plotted against temperature in the inset). It is found that under low fields ($\mu_0 H$ = 0 & 3 T), there is a slight upturn at low temperature, while higher fields ($\mu_0 H$ = 6 & 9 T) tend to suppress it. Heat capacity data measured from as-grown crystals down to ~ 0.5 K under zero applied field, revealed no signs of long-range magnetic transitions. After subtracting the approximate phonon contributions, equivalent to normalized $C_{total}/T$ of $O_2$-annealed crystals, an approximate magnetic contribution $C_{mag}/T$ for the as-grown crystals can be extracted. Then the magnetic entropy change $\Delta S_{mag}$ can be estimated to yield a small saturation value of ~ 0.13 J/mol/K which may be attributed to the weak magnetic interactions between dilute spins in the system (**Figure 5b**).

In general, we have performed extensive investigations in this work to reveal the presence of hidden hydroxides in open-air KOH flux-grown crystals of $BaNiO_3$ and how it introduces an intrinsic magnetic transition to an otherwise diamagnetic system if prepared ceramically under $O_2$ pressure. This important finding not only enriches our understanding about the Ba-Ni-O system, especially the air instability of highly oxidized $Ni^{4+}$, but also offers some insights into the underlying mechanism for the oxygen evolution reaction catalyzed by $BaNiO_3$ in alkaline media. To the best of our knowledge, only $BaNiO_3$ made in several wet-chemical routes has been reported to host this unusual catalytic behavior so far[13,14], while the analogous $ACoO_3$ (A = Ca, Sr) system made via the dry ceramic synthesis under high $O_2$ pressure can do the same trick[24]. By introducing

OH$^-$ into BaNiO$_3$, Ni$^{4+}$ (low-spin d$^6$) cations in the lattice can be partially reduced to Ni$^{3+}$ and Ni$^{2+}$, hence the originally empty $e_g$ orbitals in Ni$^{4+}$ now have the chance to be filled by electrons, which is considered a descriptor for OER activity initially proposed by Suntivich *et al*[25]. They believe the intrinsic OER activity of transition-metal oxides actually depends on the 3d electron occupancy of orbitals with an $e_g$ symmetry.[25] Thus, our experimental observations of the hidden hydroxides in the Ba-Ni-O system successfully puts this newly found OER catalyst into this category. The ceramically made ACoO$_3$ (A = Ca, Sr) compounds obey this rule as well, because Co$^{4+}$ (d$^5$) would have electrons in its $e_g$ orbitals unless it adopts a low-spin configuration, which is highly energetically unfavorable in its oxide-based materials. The more detailed catalytic mechanism may be of future interest for experts in related fields.

**Conclusions**

We have investigated open-air KOH flux-grown crystals of BaNiO$_3$, a material that in recent years has been recognized as a potentially important catalyst for the oxygen evolution reaction (OER), a vital process for diverse energy storage devices.[13,14] We performed a series of analyses for these crystals and have found a hidden story in this system. Based on our experimental observations, we conclude that the KOH flux-grown crystals made in open-air are a "wet" form of authentic BaNiO$_3$ and are unlike the material made in relatively "dry" synthetic conditions. Our work indicates that a significant portion of the O$^{2-}$ ions in the crystals grown by our method are in the form of OH$^-$ and hence necessarily reduce some of the Ni$^{4+}$ nominally present into Ni$^{3+}$ or even Ni$^{2+}$ in a disordered manner. This conclusion unifies the SC-XRD results, the XAS results, the optical characterization and the contrasting magnetism. This study serves as an excellent indication that some subtle chemical features may only be revealed by certain physical property characterization techniques, and further emphasizes the importance of bridging materials chemistry and condensed matter physics to comprehensively study solid-state materials, especially when they have significant potential for applications.

**Acknowledgements**

This research was primarily done at Princeton University, supported by the U. S. Department of Energy Division of Basic Energy Sciences, through the Institute for Quantum Matter, Grant No. DE-SC0019331. RJC would like to acknowledge discussions with Zhiwei Hu about the stability of Ni$^{4+}$ compounds. Use of the GSECARS Raman Lab System was supported by the NSF MRI Proposal (EAR-1531583). The authors acknowledge the use of Princeton's

Imaging and Analysis Center, which was partially supported by the Princeton Center for Complex Materials, an NSF-MRSEC program (No. DMR-1420541). HZW and WX were supported by U.S. DOE-BES under Contract DE-SC0023648. C.-Y. K. acknowledges the financial support from the Ministry of Science and Technology in Taiwan under grant nos. MOST 110-2112-M-A49-002-MY3 (Research Project for Newly-recruited Personnel). The authors acknowledge the support from the Max Planck-POSTECH-Hsinchu Center for Complex Phase Materials. RS and WB acknowledge the support from National Science Foundation CAREER Award No. DMR-2045760.

**Author contributions**

Lun Jin: Conceptualization, Data Curation, Formal Analysis, Investigation, Methodology, Project Administration, Software, Writing – Original Draft.

Haozhe Wang & Weiwei Xie: Data Curation, Formal Analysis, Software, Writing – Review & Editing.

Xianghan Xu, Danrui Ni & Chen Yang: Data Curation, Formal Analysis, Software, Writing – Review & Editing.

Yu-Chieh Ku, Cheng-En Liu, Chang-Yang Kuo & Chun-Fu Chang: Data Curation, Formal Analysis, Software, Writing – Review & Editing.

Raimundas Sereika & Wenli Bi: Data Curation, Formal Analysis, Software, Writing – Review & Editing.

Robert J. Cava: Conceptualization, Funding Acquisition, Project Administration, Resources, Supervision, Validation, Writing – Review & Editing.

**Competing interests**

The authors declare no competing interests.

**Supporting Information**

Supporting Information is available from the publisher or from the authors upon valid request. Correspondence and requests for materials should be addressed to L.J. and R.J.C.

- Crystallographic data, detailed structural information, diffuse reflectance spectra and oxidative TGA data of KOH flux-grown crystals; Heat capacity data of $O_2$-annealed crystals; XRD data of ceramically made $BaNiO_3$ powder.

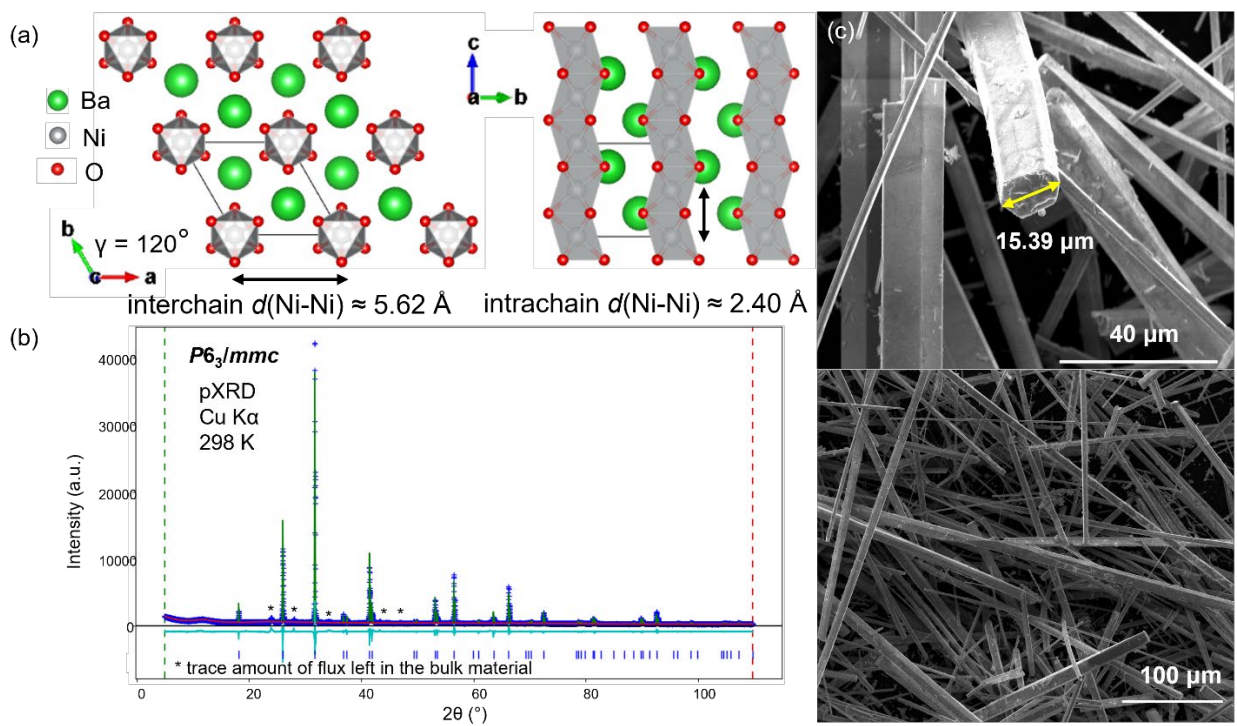

**Figure 1. Crystal Structure and morphology.** (a) SC-XRD reveals that the dark golden-green needle-shaped crystals grown from the KOH flux have the same structure as $BaNiO_3$ with no oxygen vacancies detected; (b) observed (blue), calculated (green), and difference (cyan) plots from the Rietveld refinement of crushed and ground as-grown crystals (space group $P6_3/mmc$) for the lab pXRD data collected at ambient temperature; (c) SEM images showing the morphology of as-grown crystals: hexagonal cross section area of typical size (top) and typical lengths (bottom).

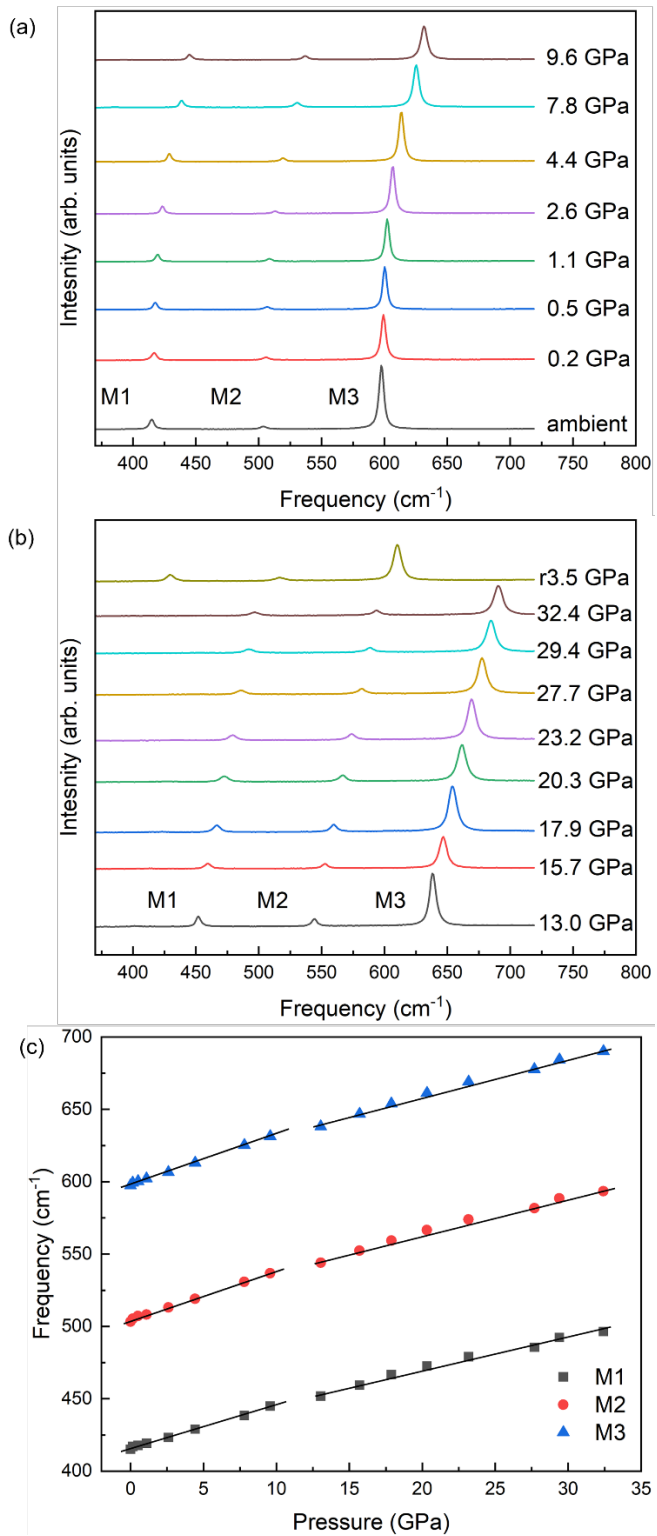

**Figure 2. High pressure Raman spectra of the as-grown crystals**. (a-b) Pressure dependence of Raman shifts up to 32.4 GPa (data marked r taken after the pressure release); (c) Evolution of Raman active modes under pressure. A slope change is observed at around 12 GPa..

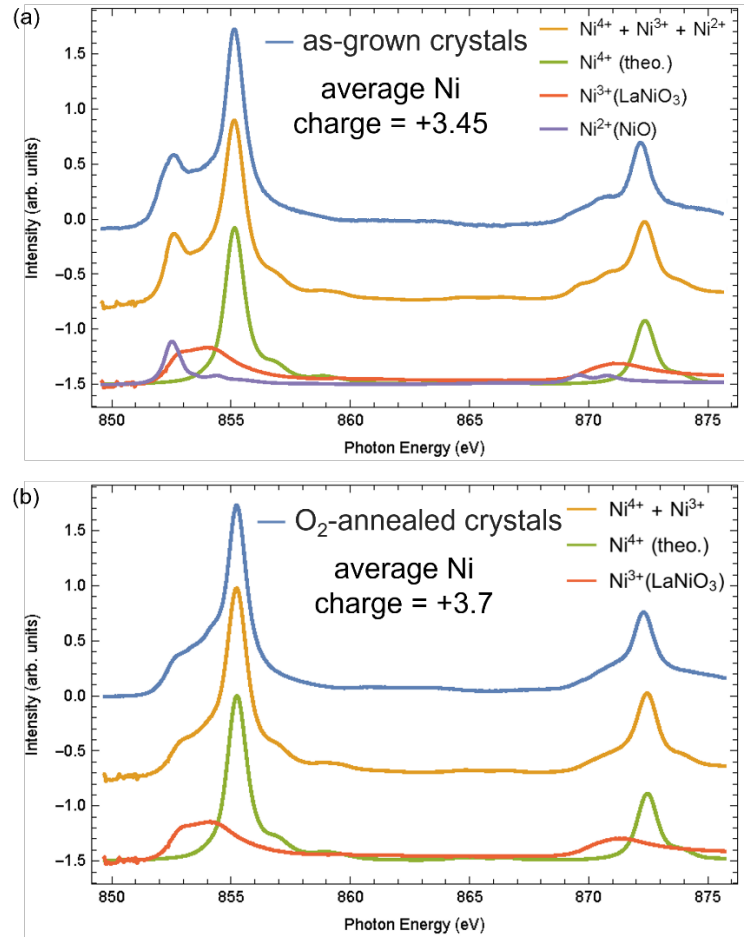

**Figure 3. XAS characterization.** The experimental Ni-$L_{2,3}$ XAS spectra of both as-grown and $O_2$-annealed crystals (b). The simulated spectrum (orange) that best matches the experimental one (blue) consists of (a) experimental NiO (purple), $LaNiO_3$ (red) and calculated $Ni^{4+}$ (green) with ratio 0.15: 0.25: 0.60 for as-grown crystals and (b) experimental $LaNiO_3$ (red) and calculated $Ni^{4+}$ (green) with ratio 0.3: 0.7 for $O_2$-annealed crystals (exposed to air for more than a month so the sample is degraded to a certain extent).

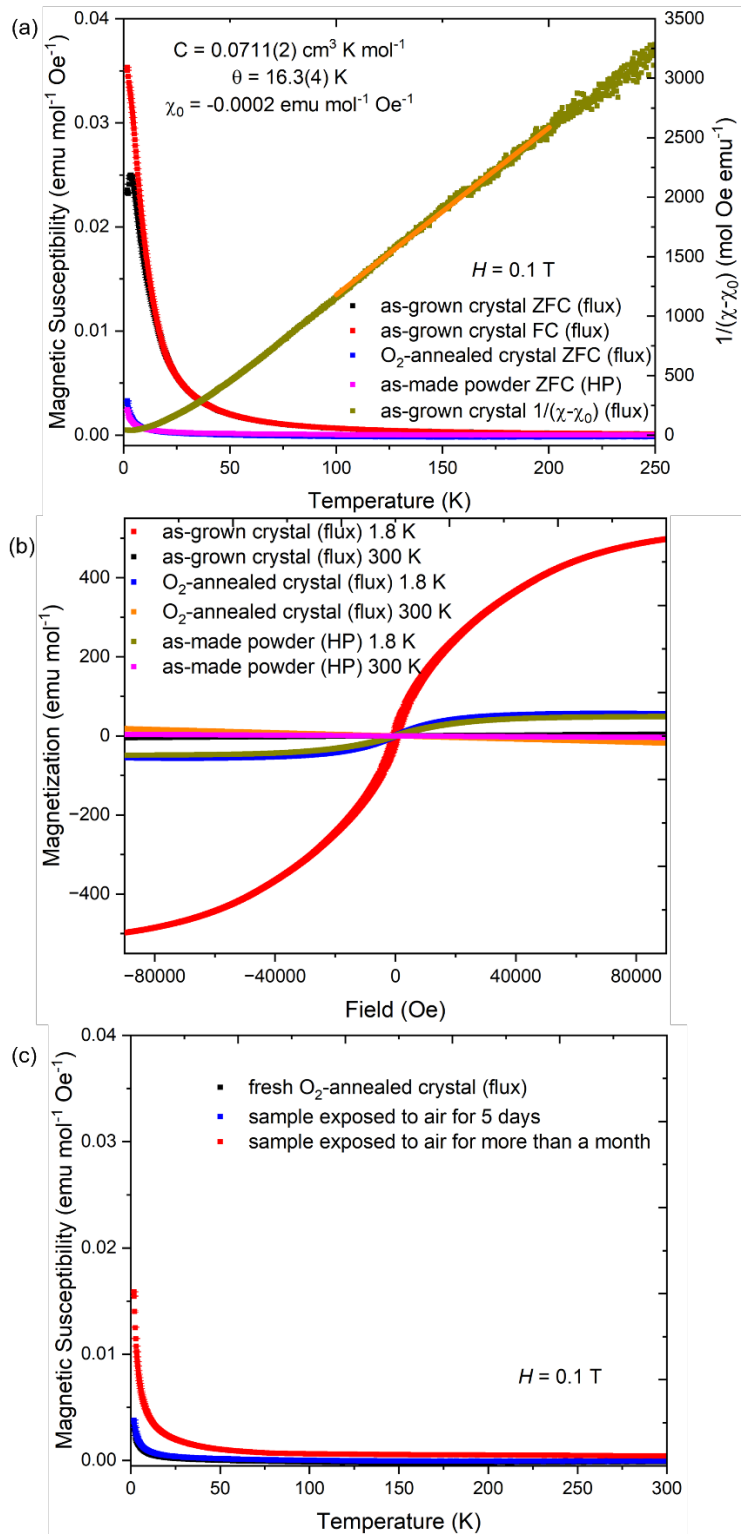

**Figure 4. Magnetic characterization.** The magnetic data collected from as-grown crystals, $O_2$-annealed crystals (fresh and degraded) and $BaNiO_3$ powder prepared via a high-pressure method.

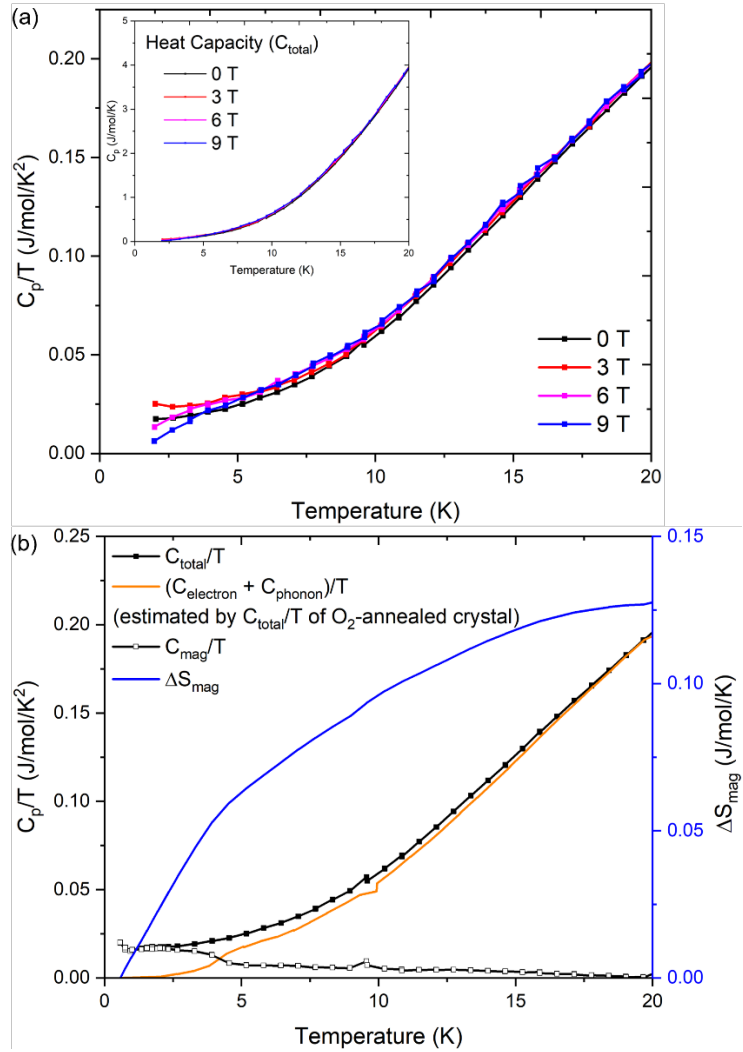

**Figure 5. Low temperature heat capacity ($C_p$) characterization.** Heat capacity data collected from as-grown crystals. (a) the resulting $C_p/T$ plotted against temperature (inset: the total heat capacity $C_{total}$ data plotted against temperature) external applied fields ($\mu_0 H$ = 0, 3, 6, 9 T); (b) All relevant $C_p/T$ plotted against temperature T to extract the magnetic contribution to heat capacity and entropy change in as-grown crystals.